\documentclass[conference,peer-reviewed]{aesconf}

\graphicspath{{./}{figures/}}

\usepackage[utf8]{inputenc}

\usepackage{microtype}

\usepackage[numbers,square]{natbib}

\usepackage{booktabs}
\usepackage{color}
\usepackage{url}

\usepackage{xcolor}
 \pagecolor{white}

\title{Ambisonics Super-Resolution Using A Waveform-Domain Neural Network}

\author{Ismael Nawfal}
\author{Symeon Delikaris Manias}
\author{Mehrez Souden}
\author{Juha Merimaa}
\author{Joshua Atkins}
\author{Elisabeth McMullin}
\author{Shadi Pirhosseinloo}
\author{Daniel Phillips}

\affil{Apple}

\correspondence{Ismael Nawfal}{inawfal@apple.com}

\lastnames{Nawfal, Ismael et al.}

\shorttitle{Ambisonics Super-Resolution Using A Waveform-Domain Neural Network}

\begin{document}

\twocolumn[
\maketitle 

\begin{onecolabstract}
Ambisonics is a spatial audio format describing a sound field. First-order Ambisonics (FOA) is a popular format comprising only four channels. This limited channel count comes at the expense of spatial accuracy. Ideally one would be able to take the efficiency of a FOA format without its limitations. We have devised a data-driven spatial audio solution that retains the efficiency of the FOA format but achieves quality that surpasses conventional renderers. Utilizing a fully convolutional time-domain audio neural network (Conv-TasNet), we created a solution that takes a FOA input and provides a higher order Ambisonics (HOA) output. This data driven approach is novel when compared to typical physics and psychoacoustic based renderers. Quantitative evaluations showed a 0.6dB average positional mean squared error difference between predicted and actual $3^\mathrm{rd}$ order HOA. The median qualitative rating showed an 80\% improvement in perceived quality over the traditional rendering approach.
\end{onecolabstract}
]

\section{Introduction}
\label{sec:intro}

Spatial audio aims at providing appropriate spatial cues to a listener to deliver a convincing, engaging, and immersive audio experience. Spatial audio can be used to preserve the original experience associated with captured sound scenes or the intentions of a content producer in the case of synthesized sound scenes. Spatial audio consists of three main components: an audio capture/synthesis block, a codec/transmission/storage block, and a rendering block. Ambisonics is a flexible spatial audio format that can be used in any or all of the above components \cite{zotter2019ambisonics,daniel2000representation}. Ambisonics can be encoded with different orders; the higher the order the higher the spatial resolution at the expense of an increased channel count and associated transmission size. 

First-order Ambisonics (FOA) is an attractive format because it comprises only four audio channels. Traditional rendering of ambisonic signals in the waveform domain is a straightforward process: the ambisonic signals are processed through a linear beamformer to provide\begin{figure}[htb]
  \centerline{\includegraphics[width=1\linewidth]{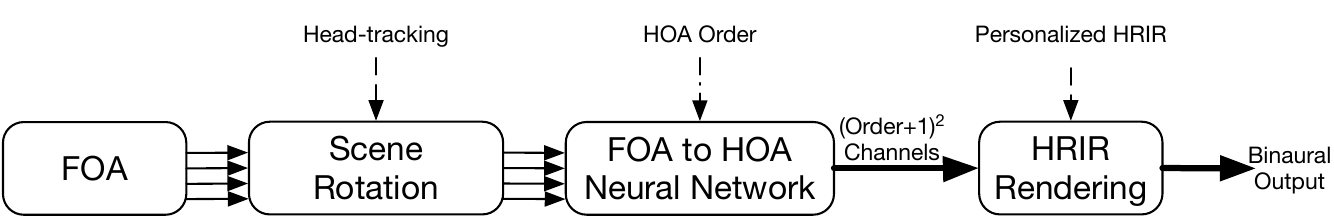}}
  \caption{Proposed method rendering binaurally via head related impulse response (HRIR). }
\label{fig:general}
\end{figure} the output signals \cite{zotter2019ambisonics}. However, FOA typically results in poor spatial audio resolution \cite{zotter2019ambisonics,pulkki2017parametric}. Numerous methods have been proposed in literature that attempt to provide a higher resolution rendering from a FOA signal \cite{pulkki2017parametric, merimaaThesis}. The current state-of-the-art FOA decoders use feature, parameter and psychoacoustic-based non-linear renderers to solve some issues associated with linear renderers, thus increasing fidelity and spatial accuracy \cite{mlfor_spatial_2022, zhu2022binaural}. Although successful in increasing spatial resolution, existing methods do not operate directly in the waveform domain. Instead, such methods rely on physical or psychoacoustic models that are prone to error when model assumptions are unfulfilled, are computationally intensive, and introduce latency \cite{pulkki2017parametric}. Higher order Ambisonics (HOA) is used when increased spatial resolution is needed but requires additional audio channels. For example, $3^\mathrm{rd}$ order HOA (HOA3) is a popular HOA format and requires 16 channels to represent a 3-D sound field. Additionally, current FOA capture devices are compact using 4 microphones while HOA capture devices require more microphones, are typically more expensive, and may only provide these higher order components within narrow frequency bandwidths\cite{zotter2019ambisonics}.

\begin{figure*}[htb]
    \centering
  \centerline{\includegraphics[width=1\linewidth]{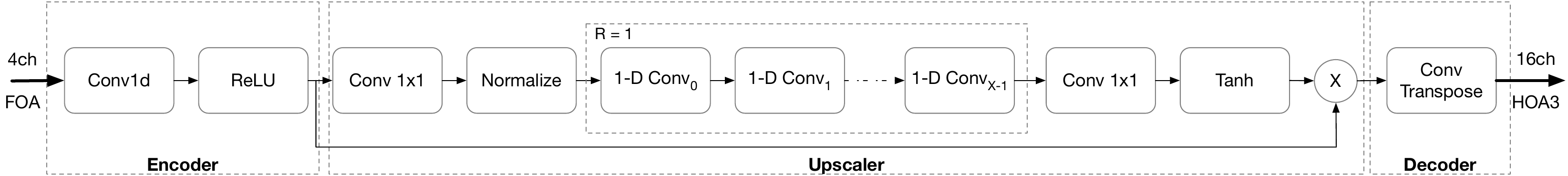}}
  \caption{The condensed Conv-TasNet topology utilized in this paper. The input is composed of the 4 channel FOA audio channels, and the output is composed of the 16 channel HOA3 components. Notably, a single repetition (R = 1) was used. 1-D Conv$_x$ is the convolutional block with dilation $x$ as introduced in \cite{luo2019conv}.}\label{fig:convtasnet}
\end{figure*}

In this study, we propose a waveform-domain, data-driven ambisonics upscaling technique. FOA signals are fed into a convolutional time-domain audio neural network (Conv-TasNet) that upscales to a HOA3 signal. The HOA3 signal can be fed to an ambisonic renderer to provide output signals for headphone or loudspeaker playback. Upscaling ambisonic signals to higher orders provides a higher resolution rendition of the captured or synthesized sound field. Upscaling FOA to HOA3 offers flexibility compared to directly rendering to headphones or a multichannel loudspeaker setup. The proposed method easily allows for head-related transfer function (HRTF) personalization for headphone rendering while also providing the flexibility of arbitrary loudspeaker placement for loudspeaker rendering. The remainder of this paper is organized as follows: Section 2 provides a background on waveform-domain Ambisonic processing, Section 3 details the proposed method, including the network architecture and data methods, Section 4 evaluates the quantitative and qualitative performance of the proposed method, and section 5 concludes. 

\section{Background}
\subsection{Waveform-domain ambisonic processing}
Conventional waveform-domain ambisonic decoders create signal-independent, linear combinations of ambisonic signals for each loudspeaker in the reproduction system. For headphone rendering, the loudspeakers can be virtualized with HRTFs \cite{ben2021binaural, hold2023magnitude, engel2021improving}. The decoding corresponds to beamforming towards the directions of the loudspeakers: each loudspeaker reproduces a signal that the corresponding beam would have captured in the recorded sound field. The spatial resolution that the beams can achieve is limited by the HOA encoding order, and optimal loudspeaker arrangements are required for best reproduction \cite{zotter2012all}.

Psychoacoustic optimizations are widely used to improve the perceived quality of FOA decoding by using different beamforming strategies at different frequencies \cite{gerzon1992,benjamin2006}. Furthermore, All-Round Ambisonic Decoding (AllRAD) was introduced to reduce the direction-dependent variation in energy and energy spread when decoding to suboptimal loudspeaker layouts \cite{zotter2012all}. However, these linear, time-invariant decoding methods cannot overcome the fundamental limitation of spatial resolution due to a limited ambisonics order. Spherical microphone array upsampling has shown promising results, allowing for the original array to be represented by higher order spherical harmonics\cite{lubeck2023spatial}. 
\begin{figure*}[htb]
    \centering
  \centerline{\includegraphics[width=1\linewidth]{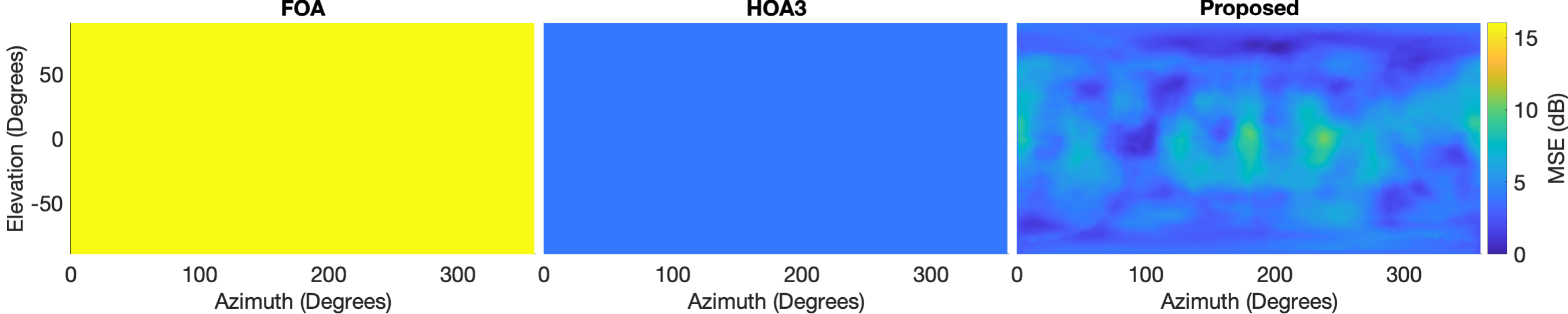}}
  \caption{Quantitative results - monophonic pink noise was rendered at 16,382 points spherical 180-design grid using FOA (Left), HOA3 (Center) and the proposed renderer (Right). Mean squared error (MSE) was calculated in decibels with respect to a point source at the same speaker position. Errors were normalized for ease of evaluation.}\label{fig:objective}
\end{figure*}
\subsection{Waveform-domain audio processing using neural networks}
Most common techniques for non-linear processing of multichannel audio are based on a time-frequency transform and an analysis-synthesis of the sound scene based on physical or psychoacoustic models. These models reflect the algorithm designer's heuristics or prior knowledge in some tractable analytical forms. Though applied with success, these models are inherently suboptimal due to tractability constraints and lack of precise knowledge of the latent data representation \cite{pulkki2017parametric}.

In the past decade, there has been a paradigm shift in audio processing, where deep learning techniques have been applied successfully to various applications, including acoustic parameter estimation \cite{bohlender2021,jukic2022}, sound field estimation \cite{chen2023sound}, noise reduction \cite{DNS2023}, speech source separation \cite{luo2019conv}, music separation \cite{hdemucs2021}, spatial capture, processing and rendering \cite{mlfor_spatial_2022}. End-to-end deep learning techniques, in particular, have been shown to be capable of accurately modeling the mapping from time-domain input to desired transformed signals. Most popular end-to-end deep learning approaches are based on fully convolutional networks, such as UNet \cite{hdemucs2021} and Conv-TasNet \cite{luo2019conv}. While UNet encodes then decodes the input signals to either denoise or separate the audio signals, Conv-TasNet mimics the conventional signal processing approaches where the signal is transformed into some learned domain then masked and transformed back to the original time domain. In this work, we investigate upscaling to HOA3, thus avoiding the issues of linear, time-invariant methods, using end-to-end learning with Conv-TasNet. In contrast to \cite{routray2019deep}, we have a fully real-time approach using convolutional neural networks.

\section{Method}
\label{sec:network}
\subsection{Network Architecture}

We address the problem of audio upscaling using Conv-TasNet \cite{luo2019conv} with some modifications. Among the many benefits of Conv-TasNet topology is its high flexibility via hyperparameters to adjust for complexity, amount of lookahead \cite{ctnlookahead2020}, and number of input and output channels. Figure \ref{fig:convtasnet} shows a simplified variant of the Conv-TasNet that was used. 

The input to our neural network is composed of FOA audio channels that are encoded in the latent space then sent through a upscaler, which essentially has the same topology as the original Conv-TasNet separator, with the exception of the output layer, which was modified to use a hyperbolic tangent. This was done to accommodate our goal of upscaling rather than masking and separating signals. Finally note that the 1-D Conv$ _x$ corresponds to the convolutional block with dilation $x$ as introduced in the original work \cite{luo2019conv}. The neural network can be adjusted to achieve different latencies and match real-time processing requirements depending on the application \cite{ctnlookahead2020}.

\subsection{Training Data and Augmentation}
Training data consisted of internal datasets comprised of 2000 hours of speech, music, and noise from diverse sources. We follow a similar dataset augmentation technique outlined in \cite{sls2023}.

Data augmentation is necessary to increase the amount of training data, reduce data scarcity, and improve model robustness and accuracy. The monophonic and anechoic training data described were augmented and encoded to represent both the source FOA and target HOA3 training data. All ambisonics encoding was done with SN3D normalization and ACN channel ordering \cite{chapman2009standard}. Augmented data were utilized for both training and validation. Offline augmentation was used primarily to aggregate data and introduce early and late room reflections. To improve the diversity of the data set, augmented sources were encoded by randomizing the following parameters:

\begin{itemize}
\item{Sources in the sound field, maximum of 12.}
\item{Location of the source in the sound field, azimuth and elevation. Distance was constant at 1 meter.}
\item{The 4 second section of audio used.}
\item{Gain applied to each source, -6dB to +6dB. }
\item{For multichannel training data, a single channel was selected for augmentation.}
\item{Simulated rooms were used to further diversify the datasets and improve the robustness of upscaling.}
\end{itemize}
\begin{figure*}[htb]
    \centering
  \centerline{\includegraphics[width=1\linewidth]{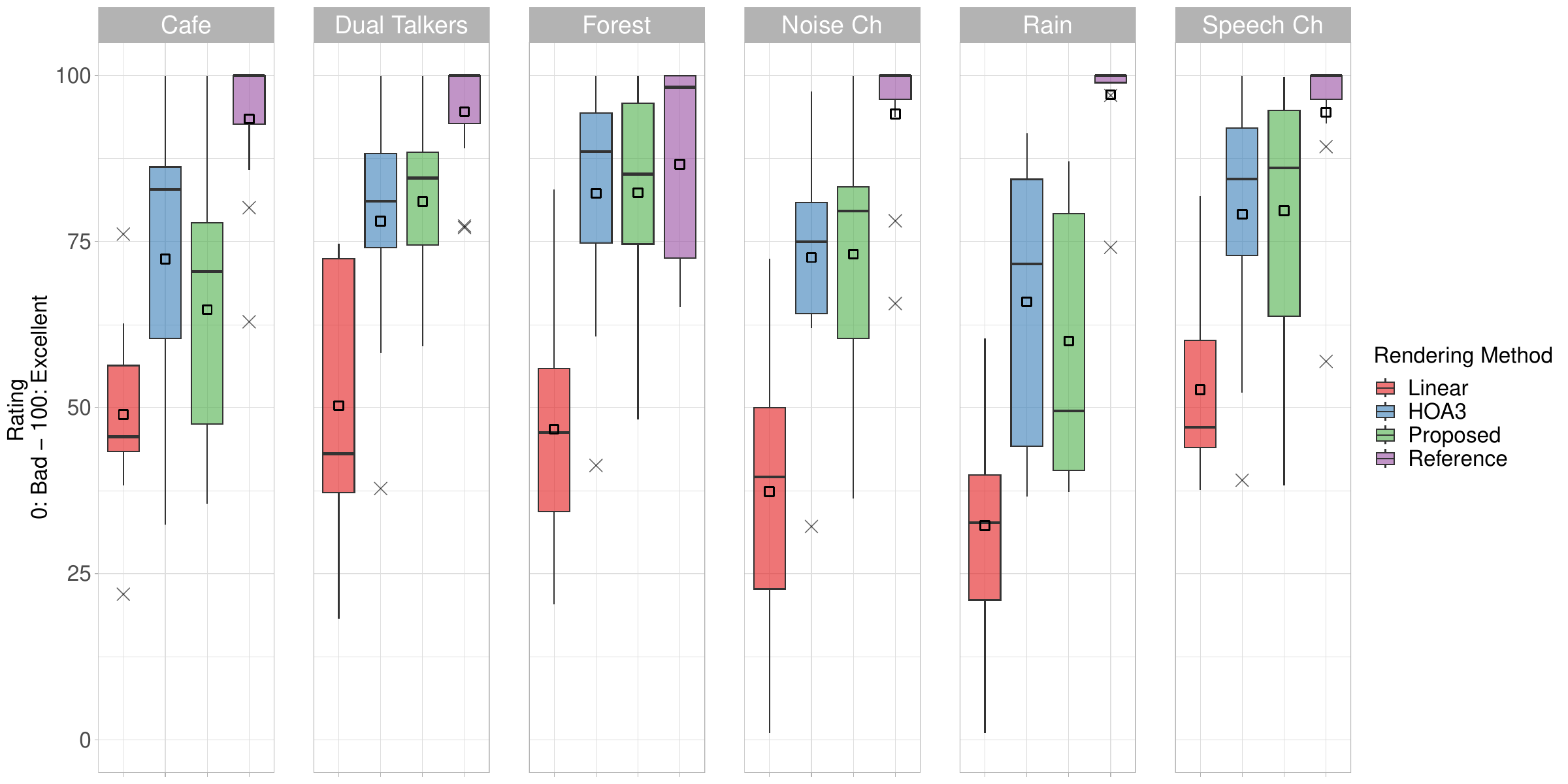}}
  \caption{Perceptual evaluation results per track - squares represent the mean rating for each renderer.}\label{fig:perceptualTrack}
\end{figure*}
\section{Evaluation}

The upscaling neural network used for both quantitative and qualitative evaluations consisted of an FOA input, HOA3 output, 384 encoder channels, and a single repetition with 256 channels resulting in 1,428,764 parameters. Training was conducted until convergence with a L1 norm reconstruction loss between target and predicted sources.

\label{sec:evaluationl}
\subsection{Quantitative Evaluation}

The proposed renderer was evaluated with a 16,382 point spherical 180-design grid \cite{an2010well} to measure positional error. For each point, a monophonic 250ms pink noise source was encoded via FOA and HOA3. The encoded FOA was then decoded via a conventional FOA decoding matrix as well as through the proposed renderer to all points on the grid. The encoded HOA3 was decoded via its corresponding conventional HOA3 decoding matrix to all points on the grid. The ground truth consisted of the same monophonic source being rendered at the desired point on the grid with no other point being excited. Mean squared error was calculated between ground truth and all three options then normalized and plotted in Figure \ref{fig:objective}.

As expected, FOA shows a significant amount of positional error when compared to the ground truth with an average positional error of 16dB, while HOA3 is significantly less at 4dB. These errors can be attributed to the smearing of energy across drivers when rendering an ambisonics source to a speaker grid. The proposed renderer using FOA as an input has a less consistent positional error around the sphere. However, the average positional error is 4.6dB, much improved over that of the FOA rendering and near to the mean error of the HOA3 rendering.

\subsection{Qualitative Evaluation} \label{subjective}

\subsubsection{Experiment Design}
The spatial audio quality of the proposed renderer was evaluated using a multiple stimuli test with a hidden reference. Listening tests were deployed using a custom web-based application, which automated the testing process, including randomizing the order of tracks and stimuli, managing results storage, and allowing users to playback and loop sections of the content. Testing was self-administered and participants set their preferred playback level prior to testing and were allowed to use headphones of their own choosing - the headphone used was not reported. Additionally, participants went through a short self-guided familiarization session which explained the rating scale and described attributes of spatial audio to take into account when rating: width, externalization, envelopment, localization precision, and naturalness.

\begin{figure}[htb]
  \centerline{\includegraphics[width=1\linewidth]{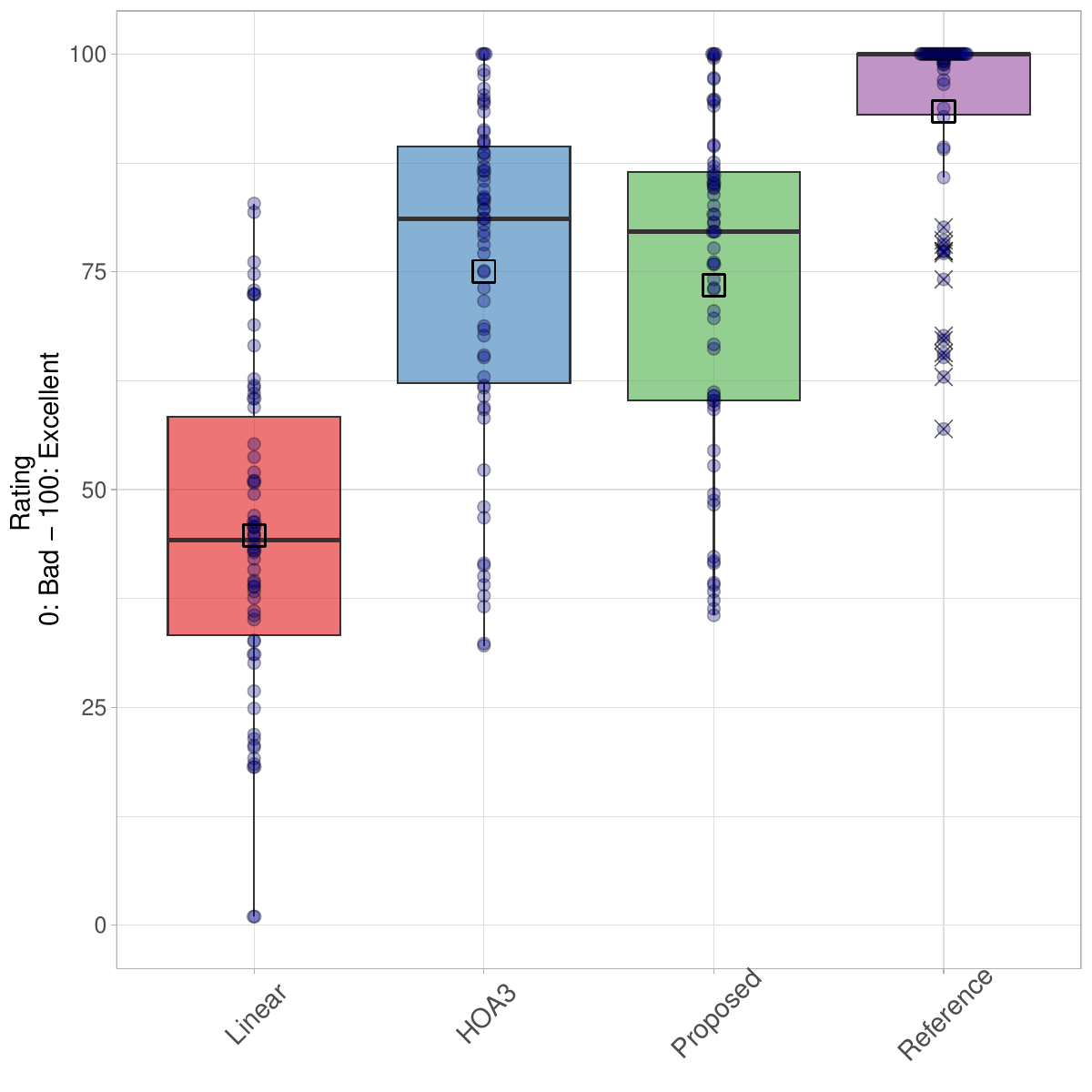}}
  \caption{Overall perceptual evaluation results - squares represent the mean rating for each renderer.}
\label{fig:perceptual}
\end{figure}

Participants were asked to rate the “Spatial Audio Quality” of the four different renderings (linearly decoded FOA, linearly decoded HOA3, the proposed renderer linearly decoded, and a hidden reference) in comparison to a known reference on a 100-point scale ranging from 0: Bad to 100: Excellent. Participants were also able to leave optional free-elicitation comments. 

They completed this task for six different 7.1.4 audio files. The audio content included a noise channel check, a male speech channel check, speech with dual talkers, a musical forest soundscape, cafe ambience, and rain. All content was pre-rendered and head-tracking was not enabled. None of the tracks were used during model training.

Thirteen participants completed testing. Two were removed from the analysis during post screening since their results demonstrated an inability to distinguish the hidden reference. All participants had experience in spatial audio evaluation and were colleagues who worked in fields related to signal processing or acoustics. 

\subsubsection{Results}
A repeated measures ANOVA was conducted with 4 renderers and 6 test cases set as within-subjects fixed factors. The data were checked for normality of the residuals using normal Q-Q plots and sphericity using a Mauchly’s Test. Based on these tests, a Greenhouse-Geisser correction was applied for the Track factor.

Renderer had a significant effect on ratings (F(3, 30) = 90.5, p $ < $ 0.001) as did Track (F(2.6, 25.7) = 3, p $ < $ 0.05). There were no significant interactions between the two. Post-hoc multiple comparison tests were run for Renderer and showed that all 4 renderings were significantly different (p $ < $ 0.01 or lower) except for HOA3 and the proposed renderer. This indicates that the upscaling from FOA to HOA3 rated similarly to the HOA3 rendering in spatial audio quality while the linear rendering rated significantly lower (as shown in Figure \ref{fig:perceptual}).

Multiple comparisons for Track indicated that there may be a significant difference (p $ < $ 0.05) in ratings between the rain and speech with dual talkers tracks shown in Figure \ref{fig:perceptualTrack}. Participants tended to rate the speech track higher than the rain track across renderings.

Five listeners left written comments about the renderings which were coded and tallied by number of repetitions and number of listeners that commented. There was little consistency across the comments for the proposed renderer. By contrast, for the linear rendering, three listeners commented it sounded narrow and had poor externalization, consistent with what has been reported in previous studies\cite{zotter2019ambisonics,pulkki2017parametric}.

\section{Conclusions}
\label{sec:conclusion}
In this paper, a novel, data-driven Ambisonics upscaling technique was proposed. The method upscales FOA to HOA3 using a convolutional time-domain audio network (Conv-TasNet). The HOA3 signals can then be rendered to any playback system, such as multichannel loudspeakers or binaural, using a linear Ambisonics decoder. The training of the network used data augmentation based on a set of 2000 hours of speech, music, and noise from diverse sources.

A quantitative evaluation showed that rendering error of the upscaled FOA was close to that of native HOA3. Qualitative evaluation further confirmed that the proposed renderer delivered equivalent spatial audio quality compared to HOA3.

Since most current Ambisonic content, as well as most commercially available Ambisonic microphones, are limited to first order, FOA is an attractive target for upscaling. We are able to upgrade existing content to HOA3. There are also potential applications for audio storage and streaming, where we can achieve a similar perceived quality with a quarter of the channels of HOA3. However, similar upscalers can be trained between any Ambisonic orders. In future work, we will explore the benefits of similar processing to even higher order content. Additionally, we will compare our proposed renderer against existing and established parametric Ambisonics decoders.

\bibliographystyle{jaes}

\bibliography{refs}

\end{document}